# De-anonymizing Social Networks

Arvind Narayanan and Vitaly Shmatikov

The University of Texas at Austin

## Abstract

*Operators of online social networks are increasingly sharing potentially sensitive information about users and their relationships with advertisers, application developers, and data-mining researchers. Privacy is typically protected by anonymization, i.e., removing names, addresses, etc.*

*We present a framework for analyzing privacy and anonymity in social networks and develop a new re-identification algorithm targeting anonymized social-network graphs. To demonstrate its effectiveness on real-world networks, we show that a third of the users who can be verified to have accounts on both Twitter, a popular microblogging service, and Flickr, an online photo-sharing site, can be re-identified in the anonymous Twitter graph with only a 12% error rate.*

*Our de-anonymization algorithm is based purely on the network topology, does not require creation of a large number of dummy "sybil" nodes, is robust to noise and all existing defenses, and works even when the overlap between the target network and the adversary's auxiliary information is small.*

## 1. Introduction

Social networks have been studied for a century [74] and are a staple of research in disciplines such as epidemiology [8], sociology [82], [12], economics [33], and many others [22], [9], [36]. The recent proliferation of online social networks such as MySpace, Facebook, Twitter, and so on has attracted attention of computer scientists, as well [45].

Even in the few online networks that are completely open, there is a disconnect between users' willingness to share information and their reaction to unintended parties viewing or using this information [16]. Most operators thus provide at least some privacy controls. Many online and virtually all offline networks (*e.g.*, telephone calls, email and instant messages, *etc.*) restrict access to the information about individual members and their relationships.

Network owners often share this information with advertising partners and other third parties. Such sharing is the foundation of the business case for many online social-network operators. Some networks are even published for research purposes. To alleviate privacy concerns, the networks are *anonymized*, *i.e.*, names and demographic information associated with individual nodes are suppressed. Such suppression is often misinterpreted as removal of "personally identifiable information" (PII), even though PII may include much more than names and identifiers (see the discussion in Appendix B). For example, the EU privacy directive defines "personal data" as "any information relating to an identified or identifiable natural person [ . . . ]; an identifiable person is one who can be identified, directly or indirectly, in particular by reference to an identification number or to one or more factors specific to his physical, physiological, mental, economic, cultural or social identity" [26].

Anonymity has been unquestioningly interpreted as equivalent to privacy in several high-profile cases of data sharing. After a New York court ruling ordering Google to hand over viewing data of over 100 million YouTube users to Viacom and the subsequent protests from privacy advocates, a revised agreement was struck under which Google would anonymize the data before handing it over [79]. The CEO of NebuAd, a U.S. company that offers targeted advertising based on browsing histories gathered from ISPs, dismissed privacy concerns by saying that "We don't have any raw data on the identifiable individual. Everything is anonymous" [18]. Phorm, a U.K. company with a similar business model, aims to collect the data on Web-surfing habits of 70% of British broadband users; the only privacy protection is that user identities are mapped to random identifiers [77]. In social networks, too, user anonymity has been used as the answer to all privacy concerns (see Section 2).

**Our contributions.** This is the first paper to demonstrate feasibility of large-scale, passive de-anonymization of real-world social networks.

First, we survey the current state of data sharing in social networks, the intended purpose of each type of sharing, the resulting privacy risks, and the wide availability of auxiliary information which can aid the attacker in de-anonymization.

Second, we formally define privacy in social networks and relate it to node anonymity. We identify several categories of attacks, differentiated by attackers' resources and auxiliary information. We also give a methodology for measuring the extent of privacy breaches in social networks, which is an interesting problem in its own right.

Third, we develop a generic re-identification algorithm for anonymized social networks. The algorithm uses only the network structure, does not make any *a priori* assumptions about membership overlap between multiple networks, and defeats all known defenses.

Fourth, we give a concrete demonstration of how our de-anonymization algorithm works by applying it to Flickr and Twitter, two large, real-world online social networks. We show that a third of the users who are verifiable members of both Flickr and Twitter[1] can be recognized in the completely anonymous Twitter graph with only 12% error rate, even though the overlap in the relationships for these members is less than 15%!

Sharing of anonymized social-network data is widespread and the auxiliary information needed for our attack is commonly available. We argue that our work calls for a substantial re-evaluation of business practices surrounding the sharing of social-network data.

## 2. State of the Union

The attacks described in this paper target anonymized, sanitized versions of social networks, using partial auxiliary information about a subset of their members. To show that both anonymized networks and auxiliary information are widely available, we survey real-world examples of social-network data sharing, most of which involve releasing *more* information than needed for our attack.

**Academic and government data-mining.** Social networks used for published data-mining research include the mobile-phone call graphs of, respectively, 7 million [63], 3 million [60], and 2.5 million [48] customers, as well as the land-line phone graph of 2.1 million Hungarian users [47]. Corporations like AT&T, whose own database of 1.9 trillion phone calls goes back decades [39], have in-house research facilities, but smaller operators must share their graphs with external researchers. Phone-call networks are also commonly used to detect illicit activity such as calling fraud [90] and for national security purposes, such as identifying the command-and-control structures of terrorist cells by their idiosyncratic sub-network topologies [39]. A number of companies sell data-mining solutions to governments for this purpose [75].

Sociologists, epidemiologists, and health-care professionals collect data about geographic, friendship, family, and sexual networks to study disease propagation and risk. For example, the Add Health dataset includes the sexual-relationship network of almost 1,000 students of an anonymous Midwestern high school as part of a detailed survey on adolescent health [2]. While the Add Health project takes a relatively enlightened stance on privacy [1], this graph has been published in an anonymized form [11].

For online social networks, the data can be collected by crawling either via an API, or "screen-scraping" (*e.g.*, Mislove *et al.* crawled Flickr, YouTube, LiveJournal, and Orkut [58]; anonymized graphs are available by request only). We stress that even when obtained from public websites, this kind of information—if publicly released—still presents privacy risks because it helps attackers who lack resources for massive crawls. In some online networks, such as LiveJournal and the Experience Project, user profiles and relationship data are public, but many users maintain pseudonymous profiles. From the attacker's perspective, this is the same as publishing the anonymized network.

**Advertising.** With the emergence of concrete evidence that social-network data makes commerce much more profitable [70], [78], network operators are increasingly sharing their graphs with advertising partners to enable better social targeting of advertisements. For example, Facebook explicitly says that users' profiles may be shared for the purpose of personalizing advertisements and promotions, as long as the individual is not explicitly identified [27]. Both Facebook and MySpace allow advertisers to use friends' profile data for ad targeting [20]. Social-network-driven advertising has been pursued by many startups [24], [59] and even Google [71], typically relying on anonymity to prevent privacy breaches [5], [25], [62].

**Third-party applications.** The number of third-party applications on Facebook alone is in the tens of thousands and rapidly growing [72]. The data from multiple applications can be aggregated and used for targeted advertising (*e.g.*, as done by SocialMedia [69]). As the notion of social networking as a feature rather than destination takes hold [4], many other networks are trying to attract application developers; on the Ning platform, which claims over 275,000 networks, each network can be considered a third-party application. The data given to third-party applications is usually not anonymized, even though most applications would be able to function on anonymized profiles [28].

Third-party applications have a poor track record of respecting privacy policies. For example, a security hole in a Facebook application developed by Slide, Inc. "exposed the birthdays, gender, and relationship status of strangers, including Facebook executives, [and] the wife of Google co-founder Larry Page" [57]. WidgetLaboratory, one of the most popular developers for the Ning platform, was banned permanently after "gathering credentials from users and otherwise creating havoc on Ning networks" [6]. Therefore, it is important to understand what a malicious third-party application can learn about members of a social network, even if it obtains the data in an anonymized form.

**Aggregation.** Aggregation of information from multiple social networks, facilitated by projects such as OpenID [64], DataPortability [21], the "social graph" project [29], and various microformats [56], potentially presents a greater threat to individual privacy than one-time data releases. Existing aggregators include FriendFeed, MyBlogLog, Jaiku (recently acquired by Google), and Plaxo; the latter even provides an open-source "social graph crawler" [67]. Aggregated networks are an excellent source of auxiliary in-formation for our attacks.

**Other data-release scenarios.** WellNet is a health-care co-ordination service which enables employers to monitor the social network in real time in order to track employees' medical and pharmacy activity [55]. The data is anonymized.

In "friend-to-friend networking," a peer-to-peer file-sharing network is overlaid on social links [68] in order to defeat censor nodes such as the RIAA. Nodes are pseudony-mous and communication is encrypted. Since traffic is typically not anonymized at the network level, the logs that can be obtained, for example, by subpoenaing the ISP are essentially anonymized social-network graphs.

Finally, consider photographs published online without identifying information. The accuracy of face recognition can be improved substantially by exploiting the fact that users who appear together in photographs are likely to be neighbors in the social network [76]. Since most online pho-tographs appear in a social-network context, they effectively represent an anonymized graph, and techniques developed in this paper can help in large-scale facial re-identification.

## 3. Related Work

**Privacy properties.** A social network consists of nodes, edges, and information associated with each node and edge. The existence of an edge between two nodes can be sen-sitive: for instance, in a sexual-relationship network with gender information attached to nodes [11] it can reveal sexual orientation. *Edge privacy* was considered in [44], [7]. In most online social networks, however, edges are public by default, and few users change the default settings [34].

While the mere presence of an edge may not be sensitive, edge attributes may reveal more information (*e.g.*, a single phone call vs. a pattern of calls indicative of a business or romantic relationship). For example, phone-call patterns of the disgraced NBA referee Tom Donaghy have been used in the investigation [91]. In online networks such as LiveJournal, there is much variability in the semantics of edge relationships [30].

The attributes attached to nodes, such as the user's inter-ests, are usually far more sensitive. Social Security numbers can be predicted from Facebook profiles with higher accu-racy than random guessing [34]; see [17] for other privacy breaches based on profile data. Even implicit attributes such as node degree can be highly sensitive, *e.g.*, in a sexual network [11]. Existing defenses focus on names and other identifiers, but basic de-anonymization only reveals that someone belongs to the network, which is hardly sensitive. As we show in the rest of this paper, however, it can be used as a vehicle for more serious attacks on privacy, including disclosure of sensitive attributes.

**De-anonymization attacks.** Backstrom *et al.* present two *active* attacks on edge privacy in anonymized social net-works [7]. These active attacks fundamentally assume that the adversary is able to modify the network prior to its re-lease: "an adversary chooses an arbitrary set of users whose privacy it wishes to violate, creates a small number of new user accounts with edges to these targeted users, and creates a pattern of links among the new accounts with the goal of making it stand out in the anonymized graph structure." Both attacks involve creating $O(\log N)$ new "sybil" nodes ($N$ is the total number of nodes), whose outgoing edges help re-identify quadratically as many existing nodes.

Active attacks are difficult to stage on a large scale. First, they are restricted to online social networks (OSNs); creating thousands of fake nodes in a phone-call or real-life network is prohibitively expensive or impossible. Even in OSNs, many operators (*e.g.*, Facebook) check the uniqueness of email addresses and deploy other methods for verifying accuracy of supplied information, making creation of a large number of dummy nodes relatively difficult.

Second, the attacker has little control over the edges *incoming* to the nodes he creates. Because most legitimate users will have no reason to link back to the sybil nodes, a subgraph with no incoming edges but many outgoing edges will stand out. As we show below, this may enable the network operator to recognize that the network has been compromised by a sybil attack. There are also other tech-niques for identifying sybil attacks in social networks [93], including methods for spammer detection deployed by OSNs that allow unidirectional edges [73].

We carried out an experiment to verify the claim that identification of subgraphs consisting primarily of sybil nodes is difficult in real-world social networks. The data for this experiment was the graph of LiveJournal obtained from Mislove *et al.* [58], crawled in late 2006. It is a directed graph with 5.3 million nodes and 77 million edges. Except for the time of the crawl, this graph is similar to that used in [7].

The cut-based attack of [7] creates 7-node subgraphs containing a Hamiltonian path. In contrast to the observation in [7] that every possible 7-node subgraph containing a Hamiltonian path occurs in the LiveJournal graph, there are no subgraphs in the LiveJournal graph that have these two properties and, furthermore, do not have any incoming edges. We conclude that active attacks are easy to detect if real users never link back to sybil nodes. More sophisticated sybil-detection techniques may work as long as only a small percentage of real users link back to sybil nodes.

The third limitation of active attacks is the fact that many OSNs require a link to be mutual before the information is made available in any form. Therefore, assuming that real users do not link back to dummy users, the links from fake nodes to real ones do not show up in the network.

We conclude that large-scale active attacks requiring cre-ation of tens of thousands of sybil nodes are unlikely to be feasible. Active attacks can still be useful in identifying or creating a small set of "seeds" to serve as a starting point for large-scale, *passive* privacy breaches. We develop such an attack in Section 5.2.

Backstrom *et al.* also describe passive attacks, in which a small coalition of users discover their location in the anonymized graph by utilizing the knowledge of the network structure around them. This attack is realistic, but again, only works on a small scale: the colluding users can only compromise the privacy of some of the users who are already their friends.

By contrast, our attack does not require creation of a large number of sybil nodes, and—as shown by our experiments on real-world online social networks—can be successfully deployed on a very large scale.

**Defenses.** Existing privacy protection mechanisms for social networks are only effective against very restricted adversaries and have been evaluated on small, simulated networks whose characteristics are different from real social networks. For example, Zheleva and Getoor give several strategies for preventing link re-identification [94], but the model ignores auxiliary information that may be available to the attacker.

An unusual attempt to prevent network operators from capitalizing on user-provided data appears in [35]. It involves scrambling the profiles when they are sent to the server and client-side unscrambling when a friend's profile is viewed. Building and running such a system involves constant reverse-engineering of communication between the client and the server. Further, all of a user's friends need to use the system, flatly contradicting the claim of incremental deployability. A similar idea appears in [52], with a more sound architecture based on a server-side Facebook application. Both approaches severely cripple social-network functionality because almost any non-trivial action other than viewing another user's profile or messages requires the server to manipulate the data in a way which is not possible under encryption.

Anonymity is a popular approach to protecting privacy. Felt and Evans propose a system where applications see randomized tokens representing users instead of actual identifiers [28]. Frikken and Golle show how to compute an anonymous graph from pieces held by different participants in order to perform privacy-preserving social-network analysis [31]. Kerschbaum and Schaad additionally enable participants to track their position in the anonymous graph [43].

Several papers proposed variants of $k$-anonymity for social networks. For example, Hay *et al.* require nodes to be automorphically equivalent [38], *i.e.*, there must exist automorphisms of the graph that map each of $k$ nodes to one another. This is an extremely strong structural requirement, which is achieved only against severely restricted adversaries: in one model, the attacker only has information about degree sequences around his target node; in another, partial knowledge of the structure in the vicinity of the target. The technique appears to work only if the average degree is low, ruling out most online social networks.

Liu and Terzi consider node re-identification assuming that the adversary's auxiliary information consists only of node degrees [51]. There is no clear motivation for this restriction. Campan and Truta propose metrics for the information loss caused by edge addition and deletion and apply $k$-anonymity to node attributes as well as neighborhood structure [15]. Zhou and Pei assume that the adversary knows the exact 1-neighborhood of the target node [95]. The anonymization algorithm attempts to make this 1-neighborhood isomorphic to $k - 1$ other 1-neighborhoods via edge addition. The experiments are performed on an undirected network with average degree 4 (an order of magnitude lower than that in real social networks) and already require increasing the number of edges by 6%. The number of edges to be added and the computational effort are likely to rise sharply with the average degree.

The fundamental problem with $k$-anonymity is that it is a syntactic property which may not provide any privacy even when satisfied (*e.g.*, if all $k$ isomorphic neighborhoods have the same value of some sensitive attributes). Crucially, all of these defenses impose arbitrary restrictions on the information available to the adversary and make arbitrary assumptions about the properties of the social network.

We argue that the auxiliary information which is likely to be available to the attacker is *global* in nature (*e.g.*, another social network with partially overlapping membership) and not restricted to the neighborhood of a single node. In the rest of this paper, we show how this information, even if very noisy, can be used for large-scale re-identification. Existing models fail to capture self-reinforcing, feedback-based attacks, in which re-identification of some nodes provides the attacker with more auxiliary information, which is then used for further re-identification. Development of a model for such attacks is our primary contribution.

# 4. Model and Definitions

## 4.1. Social network

A social network $\mathcal{S}$ consists of (1) a directed graph $G = (V, E)$, and (2) a set of attributes $\mathcal{X}$ for each node in $V$ (for instance, name, telephone number, *etc.*) and a set of attributes $\mathcal{Y}$ for each edge in $E$ (for instance, type of relationship). The model is agnostic as to whether attributes accurately reflect real-world identities or not (see Appendix C). We treat attributes as atomic values from a discrete domain; this is important for our formal definition of privacy breach (Definition 3 below). Real-valued attributes must be discretized. Where specified, we will also represent edges as attributes in $\mathcal{Y}$ taking values in $\{0, 1\}$.

In addition to the explicit attributes, some privacy policies may be concerned with implicit attributes, *i.e.*, properties of a node or an edge that are based purely on the graph structure. For example, node degree can be a sensitive implicit attribute. Implicit attributes may be leaked without disclosing any explicit attributes. For example, if the adversary re-identifies a subset of nodes in an anonymized graph, none of which are adjacent, he learns the degrees of these nodes without breaking edge privacy. Which implicit attributes should be protected depends on the specific network.

## 4.2. Data release

Our model of the data release process focuses on what types of data are released and how the data is sanitized (if at all), and abstracts away from the procedural distinctions such as whether the data is available in bulk or obtained by crawling the network. As discussed in Section 2, social-network data are routinely released to advertisers, application developers, and researchers. Advertisers are often given access to the entire graph in a (presumably) anonymized form and a limited number of relevant attributes for each node. Application developers, in current practice, get access to a subgraph via user opt-in and most or all of the attributes within this subgraph. This typically includes the identifying attributes, even if they are not essential for the application's functionality [28]. Researchers may receive the entire graph or a subgraph (up to the discretion of the network owner) and a limited set of non-identifying attributes.

"Anonymization" is modeled by publishing only a subset of attributes. Unlike naïve approaches such as $k$-anonymity, we do not distinguish identifying and non-identifying attributes (any attribute can be identifying if it happens to be known to the adversary as part of his auxiliary information). Suppressed attributes are not limited to the demographic quasi-identifiers *a priori*; we simply assume that the published attributes by themselves are insufficient for re-identification. In Section 4.4, we explain the (indirect) connection between preventing node re-identification and intuitive "privacy." In terms of entropy, most of the information in the released graph resides in the edges, and this is what our de-anonymization algorithm will exploit.

The data release process may involve perturbation or sanitization that changes the graph structure in some way to make re-identification attacks harder. As we argued in Section 3, deterministic methods that attempt to make different nodes look identical do not work on realistic networks. Other defenses are based on injecting random noise into the graph structure. The most promising one is *link prediction* [50], which produces plausible fake edges by exploiting the fact that edges in social-network graphs have a high clustering coefficient. (We stress that link prediction is far beyond the existing sanitization techniques, which mostly rely on simple removal of identifiers.) The experiments in Section 6.2 show that our algorithm is robust to injected noise, whether resulting from link prediction or not. In Appendix E, we discuss how to measure the amount of noise introduced by perturbation.

We model the data sanitization and release process as follows. First, select a subset of nodes, $V_{\mathsf{san}} \subset V$, and subsets $\mathcal{X}_{\mathsf{san}} \subseteq \mathcal{X}, \mathcal{Y}_{\mathsf{san}} \subseteq \mathcal{Y}$ of node and edge attributes to be released. Second, compute the induced subgraph on $V_{\mathsf{san}}$. For simplicity, we do not model more complex criteria for releasing edge, *e.g.*, based on edge attributes. Third, remove some edges and add fake edges. Release $S_{\mathsf{san}} = (V_{\mathsf{san}}, E_{\mathsf{san}}, \{X(v) \forall v \in V_{\mathsf{san}}, X \in \mathcal{X}_{\mathsf{san}}\}, \{Y(e) \forall e \in E_{\mathsf{san}}, Y \in \mathcal{Y}_{\mathsf{san}}\})$, *i.e.*, a sanitized subset of nodes and edges with the corresponding attributes.

## 4.3. Threat model

As described in Section 2, network owners release anonymized and possibly sanitized network graphs to commercial partners and academic researchers. Therefore, we take it for granted that the attacker will have access to such data. The main question we answer in the rest of this paper is: **can sensitive information about specific individuals be extracted from anonymized social-network graphs?**

**Attack scenarios.** Attackers fall into different categories depending on their capabilities and goals. The strongest adversary is a government-level agency interested in *global surveillance*. Such an adversary can be assumed to already have access to a large auxiliary network $S_{\mathsf{aux}}$ (see below). His objective is large-scale collection of detailed information about as many individuals as possible. This involves aggregating the anonymous network $S_{\mathsf{san}}$ with $S_{\mathsf{aux}}$ by recognizing nodes that correspond to the same individuals.

Another attack scenario involves *abusive marketing*. A commercial enterprise, especially one specializing in behavioral ad targeting [81], [92], can easily obtain an anonymized social-network graph from the network operator for advertising purposes. As described in Sections 1 and 2, anonymity is often misinterpreted as privacy. If an unethical company were able to de-anonymize the graph using publicly available data, it could engage in abusive marketing aimed at specific individuals. *Phishing and spamming* also gain from social-network de-anonymization. Using detailed information about the victim gleaned from his or her de-anonymized social-network profile, a phisher or a spammer will be able to craft a highly individualized, believable message (cf. [41]).

Yet another category of attacks involves *targeted de-anonymization* of specific individuals by stalkers, investigators, nosy colleagues, employers, or neighbors. In this scenario, the attacker has detailed contextual information about a single individual, which may include some of her attributes, a few of her social relationships, membership in other networks, and so on. The objective is to use this information to recognize the victim's node in the anonymized network and to learn sensitive information about her, including all of her social relationships in that network.

**Modeling the attacker.** We assume that in addition to the anonymized, sanitized target network $S_{\mathsf{san}}$, the attacker also has access to a *different* network $S_{\mathsf{aux}}$ whose membership partially overlaps with $S$. The assumption that the attacker possesses such an auxiliary network is very realistic. First, it may be possible to extract $S_{\mathsf{aux}}$ directly from $S$: for example, parts of some online networks can be automatically crawled, or a malicious third-party application can provide information about the subgraph of users who installed it. Second, the attacker may collude with an operator of a different network whose membership overlaps with $S$. Third, the attacker may take advantage of several ongoing aggregation projects (see Section 2). The intent of these projects is benign, but they facilitate the creation of a global auxiliary network combining bits and pieces of public information about individuals and their relationships from multiple sources. Fourth, government-level aggregators, such as intelligence and law enforcement agencies, can collect data via surveillance and court-authorized searches. Depending on the type of the attacker, the nodes of his auxiliary network may be a subset, a superset, or overlap with those of the target network.

We emphasize that even with access to a substantial auxiliary network $S_{\mathsf{aux}}$, de-anonymizing the target network $S_{\mathsf{san}}$ is a highly non-trivial task. First, the overlap between the two networks may not be large. For the entities who are members of both $S_{\mathsf{aux}}$ and $S$, some social relationships may be preserved, *i.e.*, if two nodes are connected in $S_{\mathsf{aux}}$, the corresponding nodes in $S$ are also connected with a non-negligible probability, but many of the relationships in each network are unique to that network. Even if the same entity belongs to both networks, it is not immediately clear how to *recognize* that a certain anonymous node from $S_{\mathsf{san}}$ corresponds to the same entity as a given node from $S_{\mathsf{aux}}$. Therefore, easy availability of auxiliary information does not directly imply that anonymized social networks are vulnerable to privacy breaches.

Our formal model of the attacker includes both aggregate auxiliary information (large-scale information from other data sources and social networks whose membership overlaps with the target network) and individual auxiliary information (identifiable details about a small number of individuals from the target network and possibly relationships between them). In the model, we consider edge relationship to be a binary attribute in $\mathcal{Y}$ and all edge attributes $Y \in \mathcal{Y}$ to be defined over $V^2$ instead of $E$. If $(u, v) \notin E$, then $Y[u, v] = \perp \quad \forall Y \in \mathcal{Y}$.

**Aggregate auxiliary information.** It is essential that the attacker's auxiliary information may include relationships between entities. Therefore, we model $S_{\mathsf{aux}}$ as a graph $G_{\mathsf{aux}} = \{V_{\mathsf{aux}}, E_{\mathsf{aux}}\}$ and a set of probability distributions $\mathsf{Aux}_X$ and $\mathsf{Aux}_Y$, one for each attribute of every node in $V_{\mathsf{aux}}$ and each attribute of every edge in $E_{\mathsf{aux}}$. These distributions represent the adversary's (imperfect) knowledge of the corresponding attribute value. For example, the adversary may be 80% certain that an edge between two nodes is a "friendship" and 20% that it is a mere "contact." Since we treat edges themselves as attributes, this also captures the attacker's uncertain knowledge about the existence of individual edges. This model works well in practice, although it does not capture some types of auxiliary information, such as "node $v_1$ is connected to either node $v_2$, or node $v_3$."

For an attribute $X$ of a node $v$ (respectively, attribute $Y$ of an edge $e$), we represent by $\mathsf{Aux}[X, v]$ (resp. $\mathsf{Aux}[Y, e]$) the attacker's prior probability distribution (*i.e.*, distribution given by his auxiliary information) of the attribute's value. The set $\mathsf{Aux}_X$ (resp. $\mathsf{Aux}_Y$) can be thought of as a union of $\mathsf{Aux}[X, v]$ (resp. $\mathsf{Aux}[Y, e]$) over all attributes and nodes (resp., edges).

*Aggregate auxiliary information is used in the the "propagation" stage of our de-anonymization algorithm (Section 5).*

**Individual auxiliary information (information about seeds).** We also assume that the attacker possesses detailed information about a very small[2] number of members of the target network $S$. We assume that the attacker can determine if these members are also present in his auxiliary network $S_{\mathsf{aux}}$ (*e.g.*, by matching usernames and other contextual information). The privacy question is whether this information about a handful of members of $S$ can be used, in combination with $S_{\mathsf{aux}}$, to learn sensitive information about *other* members of $S$.

It is not difficult to collect such data about a small number of nodes. If the attacker is already a user of $S$, he knows all details about his own node and its neighbors [44], [76]. Some networks permit manual access to profiles even if large-scale crawling is restricted (*e.g.*, Facebook allows viewing of information about "friends" of any member by default.) Some users may make their details public even in networks that keep them private by default. The attacker may even pay a handful of users for information about themselves and their friends [49], or learn it from compromised computers or stolen mobile phones. For example, the stored log of phone calls provides auxiliary information for de-anonymizing the phone-call graph. With an active attack (*e.g.*, [7]), the attacker may create fake nodes and edges in $S$ with features that will be easy to recognize in the anonymized version of $S$, such as a clique or an almost-clique. Since large-scale active attacks are unlikely to be feasible (see Section 3), we restrict their role to collecting individual auxiliary information as a precursor to the main, passive attack.

---

2. Negligible relative to the size of $S$. For example, in our experiments, we find that between 30 and 150 seeds are sufficient for networks with $10^5$ to $10^6$ members.

*Individual auxiliary information is used in the the "seed identification" stage of our de-anonymization algorithm (Section 5).*

## 4.4. Breaching privacy

The notion of what should be considered private varies from network to network and even from individual to individual within the network. To keep our model independent of the semantics of a particular network, we treat the *privacy policy* as a syntactic, exogenous labeling that specifies for every node attribute, edge, and edge attribute whether it should be public or private. Formally, it is a function $\mathsf{PP}: \mathcal{X} \cup \mathcal{Y} \times E \to \{\mathsf{pub}, \mathsf{priv}\}$. In Appendix D, we discuss the challenges of rigorously defining privacy policies.

In this paper, we take an "operational" approach to social-network privacy by focusing solely on node re-identification. First, it is unclear how to give a meaningful definition of social-network privacy that does not make some assumptions about the attacker's strategy and yet yields meaningful results on real-world data. Second, all currently known privacy-breaching and privacy-protection algorithms focus on node re-identification. Even edge inference, in order to be considered a meaningful privacy breach, must include learning some identifying information about the endpoints and thus implies node re-identification. Third, while anonymity is by no means sufficient for privacy[3], it is clearly necessary. A re-identification algorithm that breaks anonymity is thus guaranteed to violate any reasonable definition of privacy, as long as there are any sensitive attributes at all attached to the nodes, since the algorithm re-labels the sensitive attributes with identifying information.

We define *ground truth* to be a mapping $\mu_G$ between the nodes $V_{\mathsf{aux}}$ of the attacker's auxiliary network and the nodes $V_{\mathsf{san}}$ of the target network. Intuitively, a pair of nodes are mapped to each other if they belong to the same "entity" (see Appendix C). If $\mu_G(v)$ takes the special value $\bot$, then there is no mapping for node $v$ (*e.g.*, if $v$ was not released as part of $V_{\mathsf{san}}$). Further, $\mu_G$ need not map every node in $V_{\mathsf{san}}$. This is important because the overlap between $V_{\mathsf{san}}$ and $V_{\mathsf{aux}}$ may be relatively small. We do assume that the mapping is 1-1, *i.e.*, an entity has at most one node in each network, as discussed in Appendix C.

*Node re-identification* or re-labeling refers to finding a mapping $\mu$ between a node in $V_{\mathsf{aux}}$ and a node in $V_{\mathsf{san}}$. Intuitively, $G_{\mathsf{aux}}$ is a labeled graph and $G_{\mathsf{san}}$ is unlabeled. Node re-identification succeeds on a node $v_{\mathsf{aux}} \in V_{\mathsf{aux}}$ if $\mu(v) = \mu_G(v)$, and fails otherwise. The latter includes the case that $\mu(v) = \bot, \mu_G(v) \neq \bot$ and vice versa. Informally, re-identification is recognizing correctly that a given node in the anonymized network belongs to the same entity as a node in the attacker's auxiliary network.

*Definition 1 (Re-identification algorithm):* A node re-identification algorithm takes as input $S_{\mathsf{san}}$ and $S_{\mathsf{aux}}$ and produces a probabilistic mapping $\tilde{\mu}: V_{\mathsf{san}} \times (V_{\mathsf{aux}} \cup \{\bot\}) \to [0, 1]$, where $\tilde{\mu}(v_{\mathsf{aux}}, v_{\mathsf{san}})$ is the probability that $v_{\mathsf{aux}}$ maps to $v_{\mathsf{san}}$.

We give such an algorithm in Section 5. Observe that the algorithm outputs, for each node in $V_{\mathsf{aux}}$, a set of candidate nodes in $V_{\mathsf{san}}$ and a probability distribution over those nodes reflecting the attacker's imperfect knowledge of the re-identification mapping.

We now define the class of adversaries who attempt to breach privacy via re-identification. After constructing the mapping, the adversary updates his knowledge of the attributes of $S_{\mathsf{aux}}$ using the attribute values in $S_{\mathsf{san}}$. Specifically, he can use the probability distribution over the candidate nodes to derive a distribution over the attribute values associated with these nodes. His success is measured by the precision of his posterior knowledge of the attributes.

*Definition 2 (Mapping adversary):* A mapping adversary corresponding to a probabilistic mapping $\tilde{\mu}$ outputs a probability distribution calculated as follows:

$$\mathsf{Adv}[X, v_{\mathsf{aux}}, x] = \frac{\sum_{v \in V_{\mathsf{san}}, X[v] = x} \mu(v_{\mathsf{aux}}, v)}{\sum_{v \in V_{\mathsf{san}}, X[v] \neq \bot} \mu(v_{\mathsf{aux}}, v)}$$

$$\mathsf{Adv}[Y, u_{\mathsf{aux}}, v_{\mathsf{aux}}, y] = $$
$$\frac{\sum_{u, v \in V_{\mathsf{san}}, Y[u,v] = y} \tilde{\mu}(u_{\mathsf{aux}}, u)\tilde{\mu}(v_{\mathsf{aux}}, v)}{\sum_{u, v \in V_{\mathsf{san}}, Y[u,v] \neq \bot} \tilde{\mu}(u_{\mathsf{aux}}, u)\tilde{\mu}(v_{\mathsf{aux}}, v)}$$

Because the auxiliary graph need not be a subgraph of the target graph, the mapping may not be complete, and the mapping adversary's posterior knowledge $\mathsf{Adv}$ of an attribute value is only defined for nodes $v_{\mathsf{aux}}$ that have actually been mapped to nodes in the target graph, at least one of which has a non-null value for this attribute. Formally, $\mathsf{Adv}$ is defined if there is a non-zero number of nodes $v \in V_{\mathsf{san}}$ such that $\tilde{\mu}(v_{\mathsf{aux}}, v) > 0$ and $X[v] \neq \bot$. Edge attributes are treated similarly.

The probability of a given node having a particular attribute value can be computed in other ways, *e.g.*, by looking only at the most likely mapping. This does not make a significant difference in practice.

We say that privacy of $v_{\mathsf{san}}$ is compromised if, for some attribute $X$ which takes value $x$ in $S_{\mathsf{san}}$ and is designated as "private" by the privacy policy, the adversary's belief that $X[v_{\mathsf{aux}}] = x$ increases by more than $\delta$, which is a pre-specified privacy parameter. For simplicity, we assume that the privacy policy $\mathsf{PP}$ is global, *i.e.*, the attribute is either public, or private for all nodes (respectively, edges). More granular policies are discussed in Appendix D.

---

3. For example, suppose that the attacker can map a node in $V_{\mathsf{aux}}$ to a small set of nodes in $V_{\mathsf{san}}$ which all have the same value for some sensitive attribute. Anonymity is preserved (he does not know which of the nodes corresponds to the target node), yet he still learns the value of his target's sensitive attribute.

*Definition 3 (Privacy breach):* For nodes $u_\mathsf{aux}$, $v_\mathsf{aux} \in V_\mathsf{aux}$, let $\mu_G(u_\mathsf{aux}) = u_\mathsf{san}$ and $\mu_G(v_\mathsf{aux}) = v_\mathsf{san}$. We say that the privacy of $v_\mathsf{san}$ is breached w.r.t. adversary $\mathsf{Adv}$ and privacy parameter $\delta$ if

(a) for some attribute $X$ such that $\mathsf{PP}[X] = \mathsf{priv}$, $\mathsf{Adv}[X, v_\mathsf{san}, x] - \mathsf{Aux}[X, v_\mathsf{aux}, x] > \delta$ where $x = X[v_\mathsf{aux}]$, or

(b) for some attribute $Y$ such that $\mathsf{PP}[Y] = \mathsf{priv}$, $\mathsf{Adv}[Y, u_\mathsf{aux}, v_\mathsf{aux}, y] - \mathsf{Aux}[Y, u_\mathsf{aux}, v_\mathsf{aux}, y] > \delta$ where $y = Y[u_\mathsf{aux}, v_\mathsf{aux}]$.

Definition 3 should be viewed as a meta-definition or a template, and must be carefully adapted to each instance of the re-identification attack and each concrete attribute. This involves subjective judgment. For example, did a privacy breach occur if the attacker's confidence increased for some attributes and decreased for others? Learning common-sense knowledge from the sanitized network (for example, that all nodes have fewer than 1000 neighbors) does not intuitively constitute a privacy breach, even though it satisfies Definition 3 for the "node degree" attribute. Such common-sense knowledge must be included in the attacker's $\mathsf{Aux}$. Then learning it from the sanitized graph does not constitute a privacy breach.

## 4.5. Measuring success of an attack

While it is tempting to quantify de-anonymization of social networks in terms of the fraction of nodes affected, this results in a fairly meaningless metric. Consider the following thought experiment. Given a network $G = (V, E)$, imagine the network $G'$ consisting of $G$ augmented with $|V|$ singleton nodes. Re-identification fails on the singletons because there is no edge information associated with them, and, therefore, the naïve metric returns half the value on $G'$ as it does on $G$. Intuitively, however, the presence of singletons should not affect the performance of any de-anonymization algorithm.

This is not merely hypothetical. In many online networks, the majority of nodes show little or no observable activity after account creation. Restricting one's attention to the giant connected component does not solve the problem, either, because extraneous nodes with degree 1 instead of 0 would have essentially the same (false) impact on naïvely measured performance.

Instead, we assign a weight to each affected node in proportion to its importance in the network. Importance is a subjective notion, but can be approximated by node *centrality*, which is a well-studied concept in sociology that only recently came to the attention of computer scientists [40], [19], [54], [3], [45].

There are three groups of centrality measures: local, eigenvalue-based and distance-based. Local methods such as degree centrality consider only the neighbors of the node. Eigenvalue methods also consider the centrality of each neighbor, resulting in a convergent recursive computation. Distance-based measures consider path lengths from a node to different points in the network. A well-known eigenvalue-based measure was proposed by Bonacich in [12], while [37] presents a textbook treatment of centrality.

We find that the decision to use a centrality measure at all, as opposed to a naïve metric such as the raw fraction of nodes de-anonymized, is much more important than the actual choice of the measure. We therefore use the simplest possible measure, degree centrality, where each node is weighted in proportion to its degree. In a directed graph, we use the sum of in-degree and out-degree.

There is an additional methodological issue. For a mapped pair of nodes, should we use the centrality score from the target graph or the auxiliary graph? It is helpful to go back to the pathological example that we used to demonstrate the inadequacy of fraction-based metrics. If either of the nodes in the mapped pair is a singleton, then the de-anonymization algorithm clearly has no hope of finding that pair. Therefore, we compute the centrality in both graphs and take the minimum of the two. We believe that this formulation captures most closely the spirit of the main question we are answering in this paper: "what proportion of entities that are active in a social network and for which non-trivial auxiliary information is available can be re-identified?"

Given a probabilistic mapping $\tilde{\mu}$, we say that a (concrete) mapping is *sampled* from $\tilde{\mu}$ if for each $u$, $\mu(u)$ is sampled according to $\tilde{\mu}(u, .)$.

*Definition 4 (Success of de-anonymization):* Let $V_\mathsf{mapped} = \{v \in V_\mathsf{aux} : \mu_G(v) \neq \perp\}$. The *success rate* of a de-anonymization algorithm outputting a probabilistic mapping $\tilde{\mu}$, w.r.t. a centrality measure $\nu$, is the probability that $\mu$ sampled from $\tilde{\mu}$ maps a node $v$ to $\mu_G(v)$ if $v$ is selected according to $\nu$:

$$\frac{\sum_{v \in V_\mathsf{mapped}} \mathsf{PR}[\mu(v) = \mu_G(v)]\nu(v)}{\sum_{v \in V_\mathsf{mapped}} \nu(v)}$$

The *error rate* is the probability that $\mu$ maps a node $v$ to any node other than $\mu_G(v)$:

$$\frac{\sum_{v \in V_\mathsf{mapped}} \mathsf{PR}[\mu(v) \neq \perp \wedge \mu(v) \neq \mu_G(v)]\nu(v)}{\sum_{v \in V_\mathsf{mapped}} \nu(v)}$$

The probability is taken over the inherent randomness of the de-anonymization algorithm as well as the sampling of $\mu$ from $\tilde{\mu}$. Note that the error rate includes the possibility that $\mu_G(v) = \perp$ and $\mu(v) \neq \perp$.

The above measure only gives a lower bound on privacy breach because privacy can be violated without complete de-anonymization. Therefore, if the goal is to protect privacy, it is *not* enough to show that this measure is low. It is also necessary to show that Definition 3 is not satisfied. Observe, for example, that simply creating $k$ copies of the graph technically prevents de-anonymization and even satisfies naïve syntactic definitions such as $k$-anonymity, while completely violating any reasonable definition of privacy.

In the other direction, however, breaking Definition 4 for a large fraction of nodes—as our algorithm of Section 5 does—is sufficient to break privacy via Definition 3, as long some trivial conditions are met: at least one private attribute is released as part of $\mathcal{X}_{san}$, and the adversary possesses little or no auxiliary information about this attribute.

# 5. De-anonymization

Our re-identification algorithm runs in two stages. First, the attacker identifies a small number of "seed" nodes which are present both in the anonymous target graph and the attacker's auxiliary graph, and maps them to each other. The main, propagation stage is a self-reinforcing process in which the seed mapping is extended to new nodes using only the topology of the network, and the new mapping is fed back to the algorithm. The eventual result is a large mapping between subgraphs of the auxiliary and target networks which re-identifies all mapped nodes in the latter.

## 5.1. Seed identification

While algorithms for seed identification are not our primary technical contribution, they are a key step in enabling our overall algorithm to succeed. Here we describe one possible seed identification algorithm. The attacks in [7] can also be considered seed identification algorithms. We briefly discuss alternatives at the end of Section 6.1.

We assume that the attacker's individual auxiliary information (see Section 4.3) consists of a clique of $k$ nodes which are present both in the auxiliary and the target graphs. It is sufficient to know the degree of each of these nodes and the number of common neighbors for each pair of nodes.

The seed-finding algorithm takes as inputs (1) the target graph, (2) $k$ seed nodes in the auxiliary graph, (3) $k$ node-degree values, (4) $\binom{k}{2}$ pairs of common-neighbor counts, and (5) error parameter $\epsilon$. The algorithm searches the target graph for a unique $k$-clique with matching (within a factor of $1 \pm \epsilon$) node degrees and common-neighbor counts. If found, the algorithm maps the nodes in the clique to the corresponding nodes in the auxiliary graph; otherwise, failure is reported.

While this brute-force search is exponential in $k$, in practice this turns out not to be a problem. First, if the degree is bounded by $d$, then the complexity is $O(nd^{k-1})$. Second, the running time is heavily input-dependent, and the inputs with high running time turn out to produce a large number of matches. Terminating the algorithm as soon as more than one match is found greatly decreases the running time.

## 5.2. Propagation

The propagation algorithm takes as input two graphs $G_1 = (V_1, E_1)$ and $G_2 = (V_2, E_2)$ and a partial "seed" mapping $\mu_S$ between the two. It outputs a mapping $\mu$. One may consider probabilistic mappings, but we found it simpler to focus on deterministic 1-1 mappings $\mu \colon V_1 \to V_2$.

Intuitively, the algorithm finds new mappings using the topological structure of the network and the feedback from previously constructed mappings. It is robust to mild modifications of the topology such as those introduced by sanitization. At each iteration, the algorithm starts with the accumulated list of mapped pairs between $V_1$ and $V_2$. It picks an arbitrary unmapped node $u$ in $V_1$ and computes a score for each unmapped node $v$ in $V_2$, equal to the number of neighbors of $u$ that have been mapped to neighbors of $v$. If the strength of the match (see below) is above a threshold, the mapping between $u$ and $v$ is added to the list, and the next iteration starts. There are a few additional details and heuristics that we describe below.

**Eccentricity.** Eccentricity is a heuristic defined in [61] in the context of de-anonymizing databases. It measures how much an item in a set $X$ "stands out" from the rest, and is defined as

$$\frac{\mathsf{max}(X) - \mathsf{max}_2(X)}{\sigma(X)}$$

where $\mathsf{max}$ and $\mathsf{max}_2$ denote the highest and second highest values, respectively, and $\sigma$ denotes the standard deviation.

Our algorithm measures the eccentricity of the set of mapping scores (between a single node in $v_1$ and each unmapped node in $v_2$) and rejects the match if the eccentricity score is below a threshold.

**Edge directionality.** Recall that we are dealing with directed graphs. To compute the mapping score between a pair of nodes $u$ and $v$, the algorithm computes two scores–the first based only on the incoming edges of $u$ and $v$, and the second based only on the outgoing edges. These scores are then summed.

**Node degrees.** The mapping scores as described above are biased in favor of nodes with high degrees. To compensate for this bias, the score of each node is divided by the square root of its degree. The resemblance to cosine similarity[4] is not superficial: the rationale is the same.

**Revisiting nodes.** At the early stages of the algorithm, there are few mappings to work with, and therefore the algorithm makes more errors. As the algorithm progresses, the number of mapped nodes increases and the error rate goes down. Thus the need to revisit already mapped nodes: the mapping computed when revisiting a node may be different because of the new mappings that have become available.

---

4. The *cosine similarity measure* between two sets $X$ and $Y$ is defined when neither is empty: $\mathsf{cos}(X, Y) = \frac{|X \cap Y|}{\sqrt{|X||Y|}}$.

**Reverse match.** The algorithm is completely agnostic about the semantics of the two graphs. It does not matter whether $G_1$ is the target graph and $G_2$ is the auxiliary graph, or vice versa. Each time a node $u$ maps to $v$, the mapping scores are computed with the input graphs switched. If $v$ gets mapped back to $u$, the mapping is retained; otherwise, it is rejected.

The following pseudocode describes the algorithm in detail. `theta` is a parameter that controls the tradeoff between the yield and the accuracy.

```
function propagationStep(lgraph, rgraph, mapping)

 for lnode in lgraph.nodes:
   scores[lnode] = matchScores(lgraph, rgraph, mapping, lnode)
   if eccentricity(scores[lnode]) < theta: continue
   rnode = (pick node from right.nodes where
          scores[lnode][node] = max(scores[lnode]))

   scores[rnode] = matchScores(rgraph, lgraph, invert(mapping), rnode)
   if eccentricity(scores[rnode]) < theta: continue
   reverse_match = (pick node from lgraph.nodes where
          scores[rnode][node] = max(scores[rnode]))
   if reverse_match != lnode:
     continue

 mapping[lnode] = rnode

function matchScores(lgraph, rgraph, mapping, lnode)

 initialize scores = [0 for rnode in rgraph.nodes]

 for (lnbr, lnode) in lgraph.edges:
   if lnbr not in mapping: continue
   rnbr = mapping[lnbr]
   for (rnbr, rnode) in rgraph.edges:
     if rnode in mapping.image: continue
     scores[rnode] += 1 / rnode.in_degree ^ 0.5

 for (lnode, lnbr) in lgraph.edges:
   if lnbr not in mapping: continue
   rnbr = mapping[lnbr]
   for (rnode, rnbr) in rgraph.edges:
     if rnode in mapping.image: continue
     scores[rnode] += 1 / rnode.out_degree ^ 0.5

 return scores

function eccentricity(items)

 return (max(items) - max2(items)) / std_dev(items)

until convergence do:
 propagationStep(lgraph, rgraph, seed_mapping)
```

**Complexity.** Ignoring revisiting nodes and reverse matches, the complexity of the algorithm is $O(|E_1|d_2)$, where $d_2$ is a bound on the degree of the nodes in $V_2$. To see this, let $\mu_{part}$ be the partial mapping computed at any stage of the algorithm. For each $u \in V_1$ and each $v$ adjacent to $u$ such that $v \in \mathsf{domain}(\mu_{part})$, the algorithm examines each of the neighbors of $\mu_{part}(v)$, giving an upper bound of $|E_1|d_2$.

Assuming that a node is revisited only if the number of already-mapped neighbors of the node has increased by at least 1, we get a bound of $O(|E_1|d_1d_2)$, where $d_1$ is a bound on the degree of the nodes in $V_1$. Finally, taking reverse mappings into account, we get $O((|E_1| + |E_2|)d_1d_2)$.

## 6. Experiments

We used data from three large online social networks in our experiments. The first graph is the "follow" relationships on the Twitter microblogging service, which we crawled in late 2007. The second graph is the "contact" relationships on Flickr, a photo-sharing service, which we crawled in late 2007/early 2008. Both services have APIs that expose a mandatory username field, and optional fields name and location. The latter is represented as free-form text. The final graph is the "friend" relationships on the LiveJournal blogging service; we obtained it from the authors of [58]. The parameters of the three graphs are summarized below. In computing the average degree, the degree of a node is counted as the sum of its in- and out-degrees. Further details about the crawling process can be found in Appendix F.

| Network | Nodes | Edges | Av. Deg |
|---|---|---|---|
| Twitter | 224K | 8.5M | 37.7 |
| Flickr | 3.3M | 53M | 32.2 |
| LiveJournal | 5.3M | 77M | 29.3 |

### 6.1. Seed identification

To demonstrate feasibility of seed identification, we ran the algorithm of Section 5.1 with the LiveJournal graph as its target. Recall from Section 4.3 that the auxiliary information needed to create seed mappings comes from the users of the target network. Therefore, we can evaluate feasibility of seed identification simply by measuring how much auxiliary information is needed to identify a unique node in the target graph. We emphasize that our main de-anonymization algorithm needs only a handful of such nodes.

For simplicity, we assume that the attacker only has access to the undirected graph, where an edge is included only if it is symmetrical in the original graph. This *underestimates* the re-identification rate, because the attacker would have more information if directionality of edges were considered.

We synthetically generate auxiliary information for seed identification starting from randomly sampled *cliques*. To sample a clique of size $k$, we start from a random node and, at every stage, randomly pick a node which is adjacent to all the nodes picked so far. If there is no such node, we start over.

This method does not sample uniformly from all the cliques in the graph; the distribution of selected nodes is much more equitable. If we sample a $k$-clique uniformly, it is susceptible to anomalies in the graph that make the result meaningless. If the graph has a large clique, or even a large dense subgraph, then almost every $k$-clique sampled will belong to this large clique or subgraph.

Given a clique (specifically, a 4-clique), we assume that the attacker knows the degrees of these 4 nodes as well as the number of common neighbors of each of the 6 pairs. The auxiliary information may be imprecise, and the search algorithm treats a 4-clique in the target graph as a match as long as each degree and common-neighbor count matches within a factor of $1 \pm \epsilon$, where $\epsilon$ is the error parameter (intuitively, the higher the error, the noisier the auxiliary information and the lower the re-identification rate). Figure 1 shows how re-identification rate decreases with noise. Recall that we allow at most one match, and so the attacker never makes an error as long as his assumptions about the imprecision of his auxiliary information are correct.

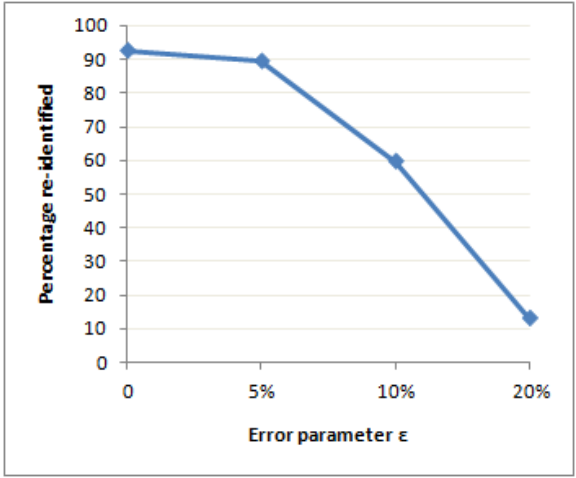

Figure 1. Seed identification

This experiment establishes that seed identification is feasible in practice. If anything, it underestimates how easy this is to do in the real world, where the attacker can use auxiliary information other than degrees and common-neighbor counts. Searching based on the structure of the target users' graph neighborhoods allows re-identification with just two or even a single node, although this is algorithmically complex.

## 6.2. Propagation

### 6.2.1. Robustness against perturbation and seed selection.
The most remarkable feature of our propagation algorithm is that it achieves "viral," self-reinforcing, large-scale re-identification regardless of the number of seeds, as long as the latter is above a (low) threshold. To study this behavior, we carried out an experiments on pairs of subgraphs, over 100,000 nodes each, of a real-world social network. In each experiment, one of the subgraphs was used as the auxiliary information, the other as the target. The graphs were artificially perturbed by adding different levels of noise to achieve various degrees of edge overlap.

**Perturbation strategy.** Given a real network graph $G = (V, E)$, our goal is to sample subsets $V_1, V_2$ of $V$ such that $V_1$ and $V_2$ have an overlap of $\alpha_V$. Overlap is measured in terms of the Jaccard Coefficient, which is defined for two sets $X$ and $Y$ if one of them is non-empty: $JC(X, Y) = \frac{|X \cap Y|}{|X \cup Y|}$. Thus, if each of two sets shares half its members with the other, the overlap is $\frac{1}{3}$. We simply partition $V$ randomly into three subsets $V_A, V_B, V_C$ of size $\frac{1-\alpha_V}{2}|V|, \alpha_V|V|, \frac{1-\alpha_V}{2}|V|$, respectively, and set $V_1 = V_A \cup V_B$ and $V_2 = V_B \cup V_C$.

We use one subgraph as the auxiliary information and the other as the anonymous target graph. As mentioned in Section 2, we believe that introducing noise via edge deletions and additions is the only realistic method of perturbing the edges. Our goal is to simulate the effect of perturbation on the target graph as follows (Procedure A):

- Derive $E'$ from $E$ by adding edges.
- Derive $E''$ from $E'$ by randomly deleting edges.
- Project $E$ and $E''$ on $V_1$ and $V_2$, respectively, to obtain $E_1$ and $E_2$.

The best way to add edges is to use link prediction, which will result in plausible fake edges. Instead of choosing a specific link prediction algorithm, we perform the following (Procedure B):

- Make two copies of $E$ and independently delete edges at random from each copy.
- Project the copies on $V_1$ and $V_2$, respectively, to get $E_1$ and $E_2$.

It should be clear that Procedure B produces more plausible edges than even the best concrete link prediction algorithm. If the link prediction algorithm is *perfect*, *i.e.*, if the edge additions accomplish the reverse of random edge deletion, then the two procedures are more or less equivalent ($E'$ in Procedure A corresponds to $E$ in Procedure B; $E$ and $E''$ in Procedure A correspond to the two perturbed copies in Procedure B). If the link prediction is not perfect, then Procedure B is better in the sense that it leads to more realistic noise, and thus makes the task of our de-anonymization algorithm harder.

This leaves the question of what fraction $\beta$ of edges to remove to get an edge overlap of $\alpha_E$. The fraction of common edges is $(1-\beta)^2$, while the fraction of edges left in at least one of the copies is $1 - \beta^2$, giving $\frac{(1-\beta)^2}{1-\beta^2} = \alpha_E$, which yields $\beta = \frac{1-\alpha_E}{1+\alpha_E}$ as the only valid solution. Note that the edge overlap is calculated for the subgraphs formed by the overlapping nodes. The overlap between $E_1$ and $E_2$ is much lower.

**Results.** We investigated the impact that the number of seeds has on the ability of the propagation algorithm to achieve large-scale re-identification, and also its robustness to perturbation.

Figure 2 shows that the selection of seeds determines whether propagation step dies out or not (cf. phase transition [89]), but whenever large-scale propagation has been achieved, the re-identification rate stays remarkably constant. We find that when the algorithm dies out, it re-identifies no more than a few dozen nodes correctly.

We performed a further experiment to study the phase transition better. A run is classified as successful if it re-identifies at least 1,000 nodes. Figure 3 shows the resulting probabilities of large-scale propagation. The phase transition is somewhat less sharp than might appear from Figure 2, although the window is almost completely in the range

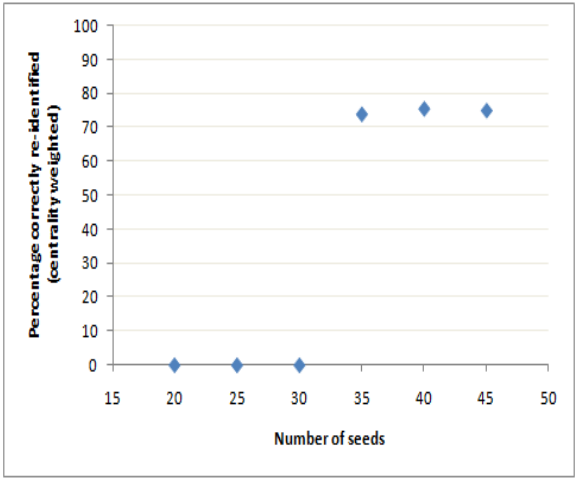

Figure 2. The fraction of nodes re-identified depends sharply on the number of seeds. Node overlap: 25%; Edge overlap: 50%

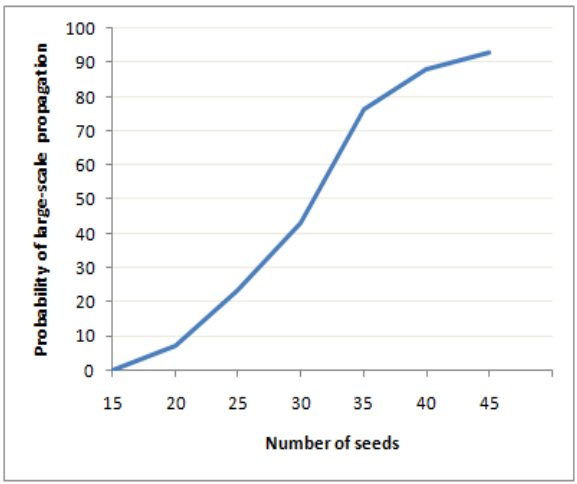

Figure 3. The phase transition in more detail. Node overlap: 25%; Edge overlap: 50%

[15,45].

It must be noted that the number of seeds required to trigger propagation depends heavily on the parameters of the graph and the algorithm used for seed selection. We therefore caution against reading too much into the numbers. What this experiment shows is that a phase transition does happen and that it is strongly dependent on the number of seeds. Therefore, the adversary can collect seed mappings incrementally until he has enough mappings to carry out large-scale re-identification.

Figure 4 shows that imprecision of the auxiliary information decreases the percentage of nodes re-identified, but cannot prevent large-scale re-identification.

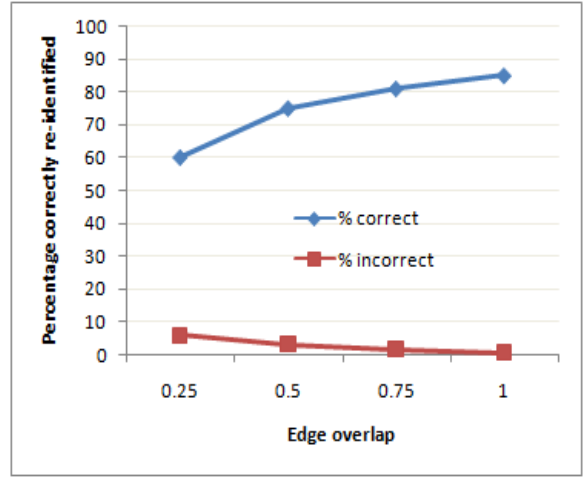

Figure 4. Effect of noise. Node overlap: 25%; Number of seeds: 50

#### 6.2.2. Mapping between two real-world social networks.

As our main experiment, we ran our propagation algorithm with the graph of Flickr as the auxiliary information and the anonymous graph of Twitter as the target.

**Ground truth.** To verify our results, we had to determine the *ground truth*, *i.e.*, the true mapping between the two graphs. We produced ground-truth mappings based on exact matches in either the username, or name field. Once a match is found, we compute a score based on a variety of heuristics on all three fields (username, name and location). If the score is too low, we reject the match as spurious.

- For usernames, we use the length to measure the likelihood that a username match is spurious. The rationale is that a username such as "tamedfalcon213" is more likely to be identifying than "joe".
- For names, we use the length of the names, as well as the frequency of occurrence of the first and last names. Rarer names indicate a stronger match.
- For locations, we use heuristics such as two-letter state abbreviations.

This resulted in around 27,000 mappings, which we will call $\mu(G)$. Since these mappings were computed with a completely different information than used by the de-anonymization algorithm, errors in the ground truth can only degrade the reported performance of our de-anonymization algorithm. We picked a random sample of the mappings and verified by human inspection that the error rate is well under 5%.

Of course, some of those who use both Flickr and Twitter may use completely different usernames and names on the two services and are thus not included in our ground-truth mappings. This has no effect on the reported performance of our algorithm. When it does recognize two nodes as belong-ing to the same user, it is rarely wrong, and, furthermore, it can successfully re-identify thousands of users.

It is possible that our algorithm has a better performance on the nodes where the ground truth is known than on other nodes. For example, users who acquire distinctive usernames on both websites might be habitual early adopters of web services. Thus, the numbers below must be interpreted with caution.

Our seed mapping consisted of 150 pairs of nodes selected randomly from $\mu(G)$, with the constraint that the degree of each mapped node in the auxiliary graph is at least 80. More opportunistic seed selection can lower the number of seeds required.

The accuracy of our algorithm on $\mu(G)$ (weighted by centrality—see Section 4.5) is summarized below:

- 30.8% of the mappings were re-identified correctly, 12.1% were identified incorrectly, and 57% were not identified.
- 41% of the incorrectly identified mappings (5% overall) were mapped to nodes which are at a distance 1 from the true mapping. It appears likely that human intelligence can be used to complete the de-anonymization in many of these cases.
- 55% of the incorrectly identified mappings (6.7% overall) were mapped to nodes where the same geographic location was reported.[5] Thus, even when re-identification does not succeed, the algorithm can often identify a node as belonging to a cluster of similar nodes, which might reveal sensitive information (recall the discussion in Section 4.4).
- The above two categories overlap; of all the incorrect mappings, only 27% (or 3.3% overall) fall into neither category and are completely erroneous.

## 7. Conclusion

The main lesson of this paper is that anonymity is not sufficient for privacy when dealing with social networks. We developed a generic re-identification algorithm and showed that it can successfully de-anonymize several thousand users in the anonymous graph of a popular microblogging service (Twitter), using a completely different social network (Flickr) as the source of auxiliary information.

Our experiments underestimate the extent of the privacy risks of anonymized social networks. The overlap between Twitter and Flickr membership at the time of our data collection was relatively small. Considering only the users who supplied their names (about a third in either network), 24% of the names associated with Twitter accounts occur in Flickr, while 5% of the names associated with Flickr accounts occur in Twitter. Since human names are not unique, this overestimates the overlap in membership. By contrast, 64% of Facebook users are also present on MySpace [66]. As social networks grow larger and include a greater fraction of the population along with their relationships, the overlap increases. Therefore, we expect that our algorithm can achieve an even greater re-identification rate on larger networks.

We demonstrated feasibility of successful re-identification based solely on the network topology and assuming that the target graph is completely anonymized. In reality, anonymized graphs are usually released with at least some attributes in their nodes and edges, making de-anonymization even easier. Furthermore, any of the thousands of third-party application developers for popular online social networks, the dozens of advertising companies, governments who have access to telephone call logs, and anyone who can compile aggregated graphs of the form described in Section 2 have access to auxiliary information which is much richer than what we used in our experiments. At the same time, an ever growing number of third parties get access to sensitive social-network data in anonymized form. These two trends appear to be headed for a collision resulting in major privacy breaches, and any potential solution would appear to necessitate a fundamental shift in business models and practices and clearer privacy laws on the subject of Personally Identifiable Information (see Appendix B).

**Acknowledgements.** The first author is grateful to Cynthia Dwork for introducing him to the problem of anonymity in social networks. Kamalika Chaudhuri deserves special thanks for collaborating on an earlier unpublished work on social network anonymity; some of the broader themes carried over to this paper. Over the last year and a half, we have had many interesting discussions with Ilya Mironov, Frank McSherry, Dan Boneh, and many others. David Molnar's help in reviewing a draft of this paper is appreciated.

This material is based upon work supported in part by the NSF grants IIS-0534198, CNS-0716158, and CNS-0746888.

---

5. This was measured by sampling 200 of the erroneous mappings and using human analysis. We consider the geographical location to be the same if it is either the same non-U.S. country, or the same U.S. state.

# Appendix A.
# Glossary

**Basic terms.**

- $S$: a social network, consisting of:
  - $G$: a graph containing nodes $V$ and edges $E$
  - $\mathcal{X}$: a set of node attributes
  - $\mathcal{Y}$: a set of edge attributes
- $X$: a node attribute, part of $\mathcal{X}$.
- $X[v]$: the value of the attribute $X$ on the node $v$
- $Y$: an edge attribute, part of $\mathcal{Y}$.
- $Y[e]$: the value of the attribute $Y$ on the edge $e$
- $\mathsf{PP}$: a privacy policy

**Sanitized and auxiliary data**

- $S_{\mathsf{san}}$: a sanitized social network, defined analogously.
- $G_{\mathsf{san}}$: a sanitized graph, containing $V_{\mathsf{san}} \subset V$ and $E_{\mathsf{san}}$, a noisy version of $E$
- $S_{\mathsf{aux}}$: the attacker's aggregate auxiliary information, consisting of
  - $G_{\mathsf{aux}} = (V_{\mathsf{aux}}, E_{\mathsf{aux}})$
  - $\mathsf{Aux} = \mathsf{Aux}_X \cup \mathsf{Aux}_Y$, (probabilistic) auxiliary information about node and edge attributes
- $\mathsf{Aux}[X, v]$: the probability distribution of the attacker's knowledge of the value of the attribute $X$ on the node $v$
- $\mathsf{Aux}[Y, e]$: likewise for edge attributes

**Re-identification**

- $\mu_G(.)$: ground truth, a 1-1 mapping between $V_{\mathsf{aux}}$ and $V_{\mathsf{san}}$
- $\tilde{\mu}(.,.)$: a probabilistic mapping output by a re-identification algorithm
- $\mu(.)$: a specific mapping between $V_{\mathsf{aux}}$ and $V_{\mathsf{san}}$ sampled from $\tilde{\mu}$
- $\nu(v)$: node centrality (Section 4.5).
- $\alpha_V$: node overlap between $V_{\mathsf{aux}}$ and $V_{\mathsf{san}}$ (Section 6.2.1)
- $\alpha_E$: edge overlap between $E_{\mathsf{aux}}$ and $E_{\mathsf{san}}$ projected on $V_{\mathsf{mapped}}$ (Section 6.2.1)
- $\epsilon$: noise parameter (for seed identification)
- $\beta$: noise parameter (for propagation; Section 6.2.1)

# Appendix B.
# On "Personally Identifiable Information"

"Personally identifiable information" is a legal term used in two related but distinct contexts. The first context is a series of breach-disclosure laws enacted in recent years in response to security breaches involving customer data that could enable identity theft.

California Senate Bill 1386 [13] is a representative example. It defines "personal information" as follows:

[An] individual's first name or first initial and last name in combination with any one or more of the following data elements, when either the name or the data elements are not encrypted:

- Social security number.
- Driver's license number or California Identification Card number.
- Account number, credit or debit card number, in combination with any required security code, access code, or password that would permit access to an individual's financial account.

Two points are worthy of note. First, the spirit of the terminology is to capture the types of information that are commonly used for authenticating an individual. This reflects the bill's intent to deter identity theft. Consequently, data such as email addresses and telephone numbers do not fall under the scope of this law. Second, it is the personal information itself that is sensitive, rather than the fact that it is possible to associate sensitive information with an identity.

The second context in which the term "personally identifiable information" appears is the privacy law. In the United States, the Privacy Act of 1974 [84] regulates the collection of personal information by government agencies, but there is no overarching law regulating private entities. At least three such acts introduced in 2005 failed to pass: the Privacy Act of 2005 [88], the Consumer Privacy Protection Act of 2005 [86], and the Online Privacy Protection Act of 2005 [87]. However, there do exist laws for specific types of data such as the Video Privacy Protection Act (VPPA) [83] and the Health Insurance Privacy and Accountability Act (HIPAA).

The language from the HIPAA Privacy Rule [85] is representative:

Individually identifiable health information is information
[. . . ]
1) That identifies the individual; or
2) With respect to which there is a reasonable basis to believe the information can be used to identify the individual.

The spirit of the law clearly encompasses deductive disclosure, and the term "reasonable basis" leaves the defining line open to interpretation by case law. We are not aware of any court decisions that define identifiability.

Individual U.S. states do have privacy protection laws that apply to any operator, such as California's Online Privacy Protection Act of 2003 [14]. Some countries other than the United States have similar generic laws, such as Canada's Personal Information Protection and Electronic Documents Act (PIPEDA) [65]. The European Union is notorious for the broad scope and strict enforcement of its privacy laws—the EU privacy directive defines "personal data" as follows [26]:

any information relating to an identified or identi-
fiable natural person [. . . ]; an identifiable person is one who can be identified, directly or indirectly, in particular by reference to an identification number or to one or more factors specific to his physical, physiological, mental, economic, cultural or social identity."

It is clear from the above that privacy law, as opposed to breach-disclosure law, in general interprets personally identifiable information broadly, in a way that is not covered by syntactic anonymization. This distinction appears to be almost universally lost on companies that collect and share personal information, as illustrated by the following Senate Committee testimony by Chris Kelly, Chief Privacy Officer of Facebook [42]:

The critical distinction that we embrace in our policies and practices, and that we want our users to understand, is between the use of personal information for advertisements in personally-identifiable form, and the use, dissemination, or sharing of information with advertisers in non-personally-identifiable form. Ad targeting that shares or sells personal information to advertisers (name, email, other contact oriented information) without user control is fundamentally different from targeting that only gives advertisers the ability to present their ads based on aggregate data.

Finally, it is important to understand that the term "personally identifiable information" has no particular technical meaning. Algorithms that can identify a user in an anonymized dataset are agnostic to the semantics of the data elements. While some data elements may be uniquely identifying on their own, *any* element can be identifying in combination with others. The feasibility of such re-identification has been amply demonstrated by the AOL privacy fiasco [10], de-anonymization of the Netflix Prize dataset [61], and the work presented in this paper. It is regrettable that the mistaken dichotomy between personally identifying and non-personally identifying attributes has crept into the technical literature in phrases such as "quasi-identifier."

# Appendix C.
# "Identity" in social networks

The correspondence between accounts or profiles (*i.e.*, network nodes) and real-world identities varies greatly from social network to social network. A wired telephone may be shared by a family or an office, while mobile phones are much more likely to belong to a single person. Some online social networks such as Facebook attempt to ensure that accounts accurately reflect real-world information [80], while others such as MySpace are notoriously lax [53]. Fake MySpace profiles have been created for pets and celebrities, and a user may create multiple profiles with contradictory or fake information.

In this paper, we eschew an explicit notion of identity and focus instead on *entities*, which are simply sources of social-network profile information that are consistent across different networks and service providers. In most cases, an entity is associated with a real-world person, but does not have to be (*e.g.*, consider a political campaign which has a YouTube account and a Twitter account). The concept of entities also allows us to capture information which is characteristic of a user across multiple networks—for example, an unusual username—but is not related to anything in the real world.

In our model, nodes are purely collections of their attributes, and to *identify* a node simply means to learn the entity to which the node belongs, whether this entity is a single person, a group, or an organization. We assume that correctly associating a node with the corresponding entity constitutes a breach of anonymity. The question of whether the entity is a single individual or not is extraneous to our model.

## Appendix D.
## Challenges of defining privacy

The fact that we are dealing with non-relational data makes it difficult to come up with a comprehensive definition of privacy in social networks. In general, one would like to say that properties of individual nodes should be privacy-sensitive and thus difficult to learn from the sanitized network, while aggregate properties should be learnable. But what counts as a "property of an individual node?" A natural candidate is any property about a $k$-neighborhood for some small $k$ (for instance, a property that a user has 3 different paths of length 2 to a known Al-Qaeda operative). Unfortunately, there does not seem to be an elegant way of choosing $k$ because social-network graphs have a very small diameter due to the "six degrees of separation" phenomenon [82].

A related approach is differential privacy [23], which in the social-network context would require that the graph look roughly the same if any single node is removed. It is not obvious how to define node removal, and far from clear how to achieve differential privacy on graph-structured data, because aggregate properties of a graph can change substantially with the removal of a single node.

Even when the privacy policy is defined as a simple labeling of attributes (as we do in Section 4.4), the policy can be *global* or *granular*. With a global policy, the same privacy label applies to a given attribute in every node (*e.g.*, email addresses are either public for all members, or private for all members). Similarly, the edges in the network are either all public, or all private. With granular policies, the privacy setting can be different for each edge and each attribute of each node.

A global policy is sufficient most of the time. In most contexts, the network operator promises users that none of their data will be released in a personally identifiable way, implying a privacy policy where all edges and all attributes are private. In other contexts, some attributes might be intuitively understood to be public (*e.g.*, node degree) and others private.

Many online social-network services such as Facebook allow users to configure their individual privacy policy with a high level of granularity. This might become a common practice in the future, but so far it appears that the vast majority of users do not change their default settings [34], [46]. There is also some ambiguity in modeling user preferences as formal privacy policies: for instance, an edge may be considered public by one endpoint and private by the other.

To keep the model simple and tractable, we do not use richer formalisms which may be suitable for some situations. For example, a multi-graph is a better model for social networks representing phone calls between individuals. We ignore the complex structure of node and edge attributes that may be relevant to privacy, such as "X knows Y through Z." We only use "public" and "private" as privacy labels, even though some networks allow more levels such as "viewable by friends," or even friends of friends.

## Appendix E.
## Measuring the effect of perturbation

The Jaccard Coefficient can be used to measure the amount of perturbation introduced to the sanitized graph $S_{\mathsf{san}}$ during the release process:

$$\frac{\sum_{u \in V_{\mathsf{san}}} \nu(u) JC(u)}{\sum_{u \in V_{\mathsf{san}}} \nu(u)}$$

where $\nu(u)$ is the centrality of the node $u$ and the Jaccard Coefficient $JC(u)$ is defined in this context as follows:

$$\frac{|\{v \in \tilde{V} : (E(u,v) \wedge \tilde{E}(u,v)) \vee (E(v,u) \wedge \tilde{E}(v,u))\}|}{|\{v \in \tilde{V} : E(u,v) \vee \tilde{E}(u,v) \vee E(v,u) \vee \tilde{E}(v,u)\}|}$$

where $\tilde{V} = V_{\mathsf{san}}$ and $\tilde{E} = E_{\mathsf{san}}$. In the above expression, the numerator counts the number of edges that are left unchanged in $E_{\mathsf{san}}$, taking directionality into account. The denominator counts all edges that exist in either direction in either $E$, or $E_{\mathsf{san}}$.

A more obvious measure that simply counts the number of edges added or removed, as a fraction of the total number of edges, would ignore the effect of perturbation on individual nodes. By contrast, our measure takes this into account, weighing nodes in proportion to their centrality in the network (this is the purpose of the $\nu$ factor).

# Appendix F.
# Notes on data acquisition

Typically, a network crawl can only recover the giant connected component. Both Twitter and Flickr allow to query only forward links. Therefore, we can expect to recover the strongly-connected component (SCC) fully and the weakly connected component (WCC) incompletely.

We crawled the entire SCC of Twitter, subject to the caveat that the Twitter API for discovering relationships is indirect; in particular, we cannot discover users whose activity on the website is "protected," *i.e.*, viewable by friends only. Interestingly, the size of the Twitter user population, at least as reflected in the connected component of regular users, turned out to be much smaller than was being reported in the media at the time of our crawl. It is also worth noting that since then Twitter has introduced crippling rate limitations on its API, which make a large-scale crawl infeasible.

We could not crawl the entire SCC of the Flickr graph due to its size. We crawled it in a priority-queue fashion, giving the highest priority to the nodes with the highest number of incoming edges from the already crawled nodes. Comparing our numbers with [58], we conclude that we have, in fact, recovered most of the SCC.

Finally, the authors of [58], who kindly provided with us with the LiveJournal data, report that their crawl covers the vast majority of the users in LiveJournal's WCC.